# A Coupled Cluster Theory Based on Quantum Electrodynamics

Sambhu N. Datta
Department of Chemistry, Indian Institute of Technology – Bombay, Powai,
Mumbai – 400 076, India
Email: sndatta@chem.iitb.ac.in

**Abstract**

An electrodynamical coupled cluster (CC) methodology starting from a covariant formalism and an equal time approximation, and finally based on the Dirac-Fock picture of the electron and positron fields and Coulomb gauge, is given here. The formalism first leads to different physical interactions from the use of an exponential cluster operator for radiative effects. Lamb, Breit and hyperfine interactions are obtained. Next, relativistic many-body effects are determined using the matter cluster in a way familiar from the nonrelativistic CC. This step can be nontrivial. By allowing the matter cluster to deviate from its traditional excitation-only form, vacuum polarization effects are generated using the pair part of Coulomb interaction. The resulting ground state correlation energy includes both relativistic and QED corrections, the latter including contributions from Lamb, Breit, hyperfine and vacuum polarization effects. The many-electron part of the theory is explicitly formulated for closed shell species. The conservatism of the second step indicates that extensions to multireference and state-specific cases are possible. To illustrate the CC approach, expressions are derived for relativistic and QED corrections to the orbital energies, configuration energies and the ground state correlation energy in a minimal basis calculation on noninteracting $H_2$ molecules. Size-consistency is maintained at every step. Because spinors of nonzero orbital angular momentum are absent, the spin-orbit interaction and Lamb shift corrections vanish in this example. However, one finds the kinetic energy correction, Darwin terms and corrections to the two-electron interaction in relativistic energy values through order $mc^2\alpha^4 Z^4$, and Breit interaction energy and hyperfine splitting of levels as QED effects through the same order. Pair energies are explicitly shown through the lowest possible orders.

## 1. Introduction

The coupled cluster (CC) treatment has developed into a mature and convenient methodology for the systematic investigation of many-body effects in atoms and molecules. Ever since its inception by Čížek and Paldus[1] and also by Bartlett[2-4], Jeziorsky and Monkhorst[5] were the first to develop the Multi-reference Coupled Cluster methods while Mukherjee and his coworkers were the first to implement a successful form of the Multi-reference Coupled Cluster[6] with a state-specific approach[7]. Bishop,[8] Farnelll et al.[9] and Kümmel[8] have discussed the coupled-cluster method, its application to and its development in physics. This method was initially developed for nuclear physics by Coester and Kümmel in the 1950s, while Čížek extended it to atomic and molecular physics in 1966. These are now standard works in many-body theory. As an almost simultaneous event, the concepts and techniques of relativistic quantum chemistry have developed into an interesting and novel subject. Several reviews and monographs have appeared, but the review by Pyykkö[11] and the book by Dyall and Fægri[12] would suffice here. Relativistic effects become pronounced in systems containing heavier atoms, and can alter the electronic structure, thereby causing measurable changes in molecular structure and energetics. For lighter atoms, intricate spectroscopic features and additional radiative effects can be observed and compared with theory. A natural outcome of these two achievements has been the development of the relativistic coupled cluster theory and the corresponding method of calculation. The relativistic CC methodology has been prepared by a straight-forward application of the coupled cluster approach familiar in the nonrelativistic theory to the solution of the relativistic wave equation based on a relativistic Hamiltonian such as the Dirac-Coulomb-Breit operator in its projected form. The latter operator involves interactions that are in principle phenomenological. It can be best described as the Hamiltonian of the field theory of matter, and of course it can be derived from quantum electrodynamics (QED).

The presently known formulation of relativistic CC theory has several attractive features. (1) It is normally based on the Dirac-Fock orbitals that can be determined either from the Dirac-Coulomb Hamiltonian[13-14] or from the Dirac-Coulomb-Breit Hamiltonian[15]. In both cases, Breit interaction energy is obtained as expectation value over the Dirac-Fock ground state wave function.[13-20] In the second case, it also contributes to the determination of the ground state configuration through the SCF process. (2) Safety from variation collapse is generally achieved at the Dirac-Fock level by using the matrix representation of operators as suggested by Grant and his coworkers.[21-22] (3) For a finite basis calculation, spurious spinors of negative energy are not taken into account, in order to avoid continuum dissolution. One must use projected interactions. Furthermore, the use of numerical Dirac-Fock orbitals can account for a proper projection.[18] (4) A multi-reference coupled cluster treatment has also been formulated.[20] (5) Some authors like to base the relativistic CC on the Douglas-Kroll-Hess transformation and use the two-component spinors in order to bypass the two theoretical problems mentioned in (2) and (3).[23] This comes at the cost that the calculation remains approximate through any finite order, a large basis set is required, and the convergence is often slow. Besides, the evaluation of radiative effects becomes tedious. However, everything has not been so rosy. Sometimes, in their haste, authors may neglect the spin-orbit splitting of orbitals while selecting the basis spinors. A relativistic treatment is for numerical accuracy that costs computational time, space and effort. Saue et al.[24] utter the caveat that the effects of the spin-orbit interaction should be fully retained in the treatment.

It stands logical to work out a relativistic CC method that is based on the Hamiltonian of QED rather than starting at the halfway mark. This task is accomplished in the present work. The first step involves the derivation of familiar interactions, namely, Lamb shift interaction, Breit interaction (a combination of Gaunt and retarded interactions) and hyperfine interactions from a radiative cluster approach in QED. It is then possible to get the Dirac-Fock (DF) ground state energy and Lamb, Breit and hyperfine corrections to it. The correlation energy and the correlated wave function are obtained from the second step using the matter cluster. Indeed one obtains the relativistic correlation energy along with possible Lamb, Breit and hyperfine corrections. As part of the second step, the matter cluster is allowed to deviate from conventionality so that the Coulombic pair terms give rise to energy corrections due to the creation and annihilation of virtual electron-positron pairs. Lamb shift can be observed and Breit interaction energies can be estimated for atoms. However, the internal motions of a general molecule often mask them from being detected. The hyperfine splitting in the ground state can be easily estimated, and it is detectable even for a molecule from magnetic resonance spectroscopies. This work provides a justification for the relativistic CC methodologies that have been already developed, and introduces additional interactions to the theoretical treatment.

## 2. Theoretical Background

A field theoretical formulation needs to start from the choice of a specific physical picture such as the free particle picture or the Furry bound state interaction picture. The mean field picture is adopted here, with $u_{\sigma m}$ ($v_{\sigma n}$) being the positive-energy (negative-mass) eigenspinors of the $N$-electron relativistic Fock operator

$$F = h_{D,ext} + \frac{1}{2} \upsilon_{DF}(\boldsymbol{r}),$$

$$\upsilon_{DF}(\boldsymbol{r}) = \sum_{u_{\sigma m}}^{occ} [\hat{J}_{\sigma m}(\boldsymbol{r}) - \hat{K}_{\sigma m}(\boldsymbol{r})],$$

(1)

where the external-field Dirac operator is given by

$$h_{D,ext} = \beta mc^2 + c\boldsymbol{\alpha} \cdot \boldsymbol{p} + eA^0_{ext}(\boldsymbol{r}).$$  (2)

The corresponding matter field is written as

$$\psi(\boldsymbol{r},t) = \psi_+(\boldsymbol{r},t) + \psi_-^\dagger(\boldsymbol{r},t)$$

where $\psi_+$ is the field operator for the bound and scattered states of positive energy for the attractive interaction between the particle and nuclear centers, and $\psi_-$ is the operator for the scattered states of the positive-energy antiparticle, (charge conjugated to the eigenstates of the negative-energy electron). In diagonal representation,

$$\psi_+(\boldsymbol{r},t) = \underset{m\sigma}{\mathrm{S}} a_{m\sigma}(t) u_{m\sigma}(\boldsymbol{r}),$$

$$\psi_-(\boldsymbol{r},t) = \underset{p\sigma}{\mathrm{S}} b_{p\sigma}(t) v_{p\sigma}^*(\boldsymbol{r}).$$

(3)

where $a, a^\dagger$ (and $b, b^\dagger$) are particle (and antiparticle) destruction and creation operators. The fermionic field operators obey the equal time anticommutation rule,

$$[\psi_\pm(\boldsymbol{r},t), \psi_\pm^\dagger(\boldsymbol{r}',t)]_+ = \delta^3(\boldsymbol{r} - \boldsymbol{r}'),$$

$$[\psi_\pm(\boldsymbol{r},t), \psi_\pm(\boldsymbol{r}',t)]_+ = 0.$$

(4)

The field operators are used to write down different components of Hamiltonian. To begin, the external-field electronic Hamiltonian operator is written as

$$\hat{H}_{D,ext} = \int d^3r : \psi^\dagger(\mathbf{r}) h_{D,ext} \psi(\mathbf{r}) : \tag{5}$$

and the interparticle Coulomb interaction as

$$\hat{H}_C = \frac{e^2}{2} \int d^3r_1 \int d^3r_2 \, \hat{\rho}_D(\mathbf{r}_1) \frac{1}{|\mathbf{r}_1 - \mathbf{r}_2|} \hat{\rho}_D(\mathbf{r}_2) \tag{6}$$

where $\hat{\rho}_D(\mathbf{r})$ is the field-theoretical density operator,

$$\hat{\rho}_D(\mathbf{r}) = :\psi^\dagger(\mathbf{r})\psi(\mathbf{r}): . \tag{7}$$

Another interesting quantity is the probability current of field theory,

$$\hat{J}_D(\mathbf{r}) = c :\psi^\dagger(\mathbf{r}) \boldsymbol{\alpha} \psi(\mathbf{r}): . \tag{8}$$

The 4-currents are to be used in writing down interaction between radiation and matter. Sucher[25] asked for utilizing only the part of Coulomb interaction that is projected onto the positive-energy subspace so that the continuum dissolution problem can be avoided.

The Dirac-Coulomb Hamiltonian (in coordinate representation here) gives rise to the Hamiltonian operator of quantum field theory (QFT),

$$\hat{H}_{QFT} = \hat{H}_{D,ext} + \hat{H}_C . \tag{9}$$

One may carry out a mean-field treatment with this operator. The electronic ground state configuration in this picture is represented by the state vector $|\Psi_N^0\rangle$ and the excited states configurations are written as $|\Psi_N^n\rangle$ (for $n \neq 0$). These vectors are confined to the $N$-electron sector of Fock space:

$$\begin{aligned}
&\Lambda_+ |\Psi_N^n\rangle = |\Psi_N^n\rangle, \\
&\Lambda_+ = \prod_{i=1}^N \lambda_+(i), \qquad \lambda_+(i) = \underset{\sigma m}{S} |\sigma m(i)\rangle\langle \sigma m(i)|.
\end{aligned} \tag{10}$$

The projected interaction is written as $\Lambda_+ \hat{H}_C \Lambda_+$. The projected Hamiltonian of QFT, $\hat{H}_{QFT}(projected) = \hat{H}_{D,ext} + \Lambda_+ \hat{H}_C \Lambda_+$, is the configuration-space equivalent of the QFT no-pair Hamiltonian restricted to the $N$-electron sector of Fock space.[25] The corresponding DF ground configuration energy is

$$\begin{aligned}
E_N^0 &= \langle \Psi_N^0 | \hat{H}_{QFT}(projected) | \Psi_N^0 \rangle \\
&= \langle \Psi_N^0 | \hat{H}_{D,ext} | \Psi_N^0 \rangle + \langle \Psi_N^0 | \Lambda_+ \hat{H}_C \Lambda_+ | \Psi_N^0 \rangle.
\end{aligned} \tag{11}$$

Because of the restriction (10) on the state vector, the pair part of Coulomb interaction $\hat{H}_C^{Pair} = \hat{H}_C - \hat{H}_C^{no-pair} = \hat{H}_C - \Lambda_+ \hat{H}_C \Lambda_+$ makes zero contribution to the energy of the DF ground state configuration.

The Dirac-Fock Hamiltonian is a sum of the Fock operators for all $N$ electrons, explicitly written as

$$\begin{aligned}
&\hat{H}_{DF} = \sum_{i=1}^N F(i) = \hat{H}_{D,ext} + \hat{V}_{DF}, \\
&\hat{V}_{DF} = \int d^3r : \psi^\dagger(\mathbf{r}) \upsilon_{DF}(\mathbf{r}) \psi(\mathbf{r}) : .
\end{aligned} \tag{12}$$

The eigenvalues of the Dirac-Fock Hamiltonian operator are written as $E_{DF}^n$,

$$\hat{H}_{DF} |\Psi_N^n\rangle = E_{DF}^n |\Psi_N^n\rangle . \tag{13}$$

One may notice that the DF ground state energy $E^0_{DF}$ differs from the energy of the DF ground configuration $E^0_N$ by an additional amount of the no-pair interaction energy.

The quantized radiation field is prescribed now. Operators for the creation and destruction of a photon are written as $A^\dagger_{k\lambda}$ and $A_{k\lambda}$ where $k$ is the wave vector while $\lambda$ is the unit vector in one of the two directions of transverse polarization. These operators follow the Bose particle commutation rules $[A_{k\lambda}, A_{k'\lambda'}]_- = 0$ and $[A_{k\lambda}, A^\dagger_{k'\lambda'}]_- = \delta_{k,k'}\delta_{\lambda,\lambda'}$. Furthermore, $\omega_k = ck$, $\hat{N}_{k\lambda} = A^\dagger_{k\lambda} A_{k\lambda}$ is the number operator and $|N_{k\lambda}\rangle$ is the number state such that
$$\langle N_{k\lambda} | N_{k'\lambda'} \rangle = \delta_{k,k'}\delta_{\lambda,\lambda'}, \tag{14}$$
while
$$\begin{aligned}
A_{k\lambda} | N_{k'\lambda'} \rangle &= \delta_{k,k'}\delta_{\lambda,\lambda'} (2\pi\hbar c^2/\omega_k)^{1/2} (N_{k\lambda})^{1/2} | N_{k\lambda} - 1 \rangle, \\
A^\dagger_{k\lambda} | N_{k'\lambda'} \rangle &= \delta_{k,k'}\delta_{\lambda,\lambda'} (2\pi\hbar c^2/\omega_k)^{1/2} (N_{k\lambda}+1)^{1/2} | N_{k\lambda} + 1 \rangle,
\end{aligned} \tag{15}$$
and $\langle \hat{N}_{k\lambda} \rangle = (2\pi\hbar c/k) N_{k\lambda}$. In transverse gauge, the electromagnetic 4-potential operators are
$$\begin{aligned}
A^0(r,t) &= \int d^3r' \frac{\hat{\rho}^0_D(r',t)}{|r-r'|}, \\
A(r,t) &= \frac{1}{\sqrt{\Omega}} \sum_{k\lambda} \left[ A_{k\lambda} \lambda e^{i(k \cdot r - \omega_k t)} + h.c. \right].
\end{aligned} \tag{16}$$
where $\Omega$ is the volume in which the photons are counted. The Hamiltonian for the quantized radiation field per unit volume is
$$\hat{H}^0_{rad} = \Omega^{-1} \sum_{k\lambda} (k/2\pi\hbar c) \hat{N}_{k\lambda} \hbar\omega_k \tag{17}$$
and the state vector is $|\Psi^0_{rad}\rangle = |\{N_{k\lambda}\}\rangle$ such that the energy density is
$$E^0_{rad} = \Omega^{-1} \sum_{k\lambda} N_{k\lambda} \hbar\omega_k. \tag{18}$$

The covariant interaction of the matter 4-current with the radiation 4-potential would be
$$\begin{aligned}
\hat{H}_{int} &= e \int d^3r \, J_\mu A^\mu \\
&= \hat{H}^{(1)}_{int} + \hat{H}^{(2)}_{int}
\end{aligned} \tag{19}$$
where the interaction involving a longitudinal photon and a transverse virtual photon between two species are
$$\hat{H}^{(1)}_{int} = e^2 \int d^3r \, \hat{\rho}_D(r,t) A^0(r,t) \tag{20}$$
and
$$\hat{H}^{(2)}_{int} = -\frac{e}{c} \int d^3r \, \hat{J}_D(r,t) \cdot A(r,t), \tag{21}$$
respectively. In Coulomb gauge, $eA^0(r,t) = \int d^3r' \frac{e\hat{\rho}_D(r',t)}{|r-r'|}$. However, all the pair interactions have been double counted in $\hat{H}^{(1)}_{int}$. One needs to retain only one interaction between each electron pair, that is, only $\hat{H}_C = \hat{H}^{(1)}_{int}/2$, thereafter causing a departure from an approximate covariance. The vector interactions embodied in $\hat{H}^{(2)}_{int}$ are responsible for known QED effects such as Breit interaction of order $mc^2\alpha^4 Z^2$ (sum of the electron-electron magnetic interaction or Gaunt term and the retarded interaction), Lamb shift of order $mc^2\alpha^5Z^4$ (a part of the electron self-energy), and hyperfine interaction of order $(m^2/M)c^2\alpha^4Z^3$ (magnetic interaction between the electron and nucleus). The quantity $\alpha$ is fine structure constant. One writes the Hamiltonian operator of QED as

$$\hat{H}_{QED} = \hat{H}^0_{rad} + \hat{H}_{D,ext} + \hat{H}_C + \hat{H}^{(2)}_{int}, \quad (\hat{H}_C = \Lambda_+ \hat{H}_C \Lambda_+ + \hat{H}^{Pair}_C). \tag{22}$$

This Hamiltonian is used in the present work.

## 3. Relativistic Coupled Cluster Theory

Current relativistic CC formalisms are based on $\hat{H}_{QFT}$. There are two slightly different methods. In both cases the correlation corrections are calculated using the purely electronic cluster operators. In the first and makeshift form, the mean field is derived only from the (projected) Coulomb term, and a nonrelativistic-like CC is carried out. One may also consider Breit interaction along with the Hamiltonian of QFT. Breit interaction energy can be obtained either as an expectation value over the HF ground state configuration, or as an expectation value over the wave function that results from the CC treatment. The difference between the two would represent Breit interaction correction to correlation energy. In the second and refined version, the mean field is determined by both (projected) Coulomb and (projected) Breit interactions, and subsequently the correlation corrections are calculated. In this case the orbitals (Dirac-Fock-Breit spinor eigenvectors) as well as the correlated wave function (written using the coefficients of the cluster operators) are influenced by Breit operator. The total correlation energy here would be slightly different from the sum of the Coulomb correlation and the Breit term induced correction to it. These two approaches, makeshift and refined, have been adopted in earlier work as found in references 13-15, 17-20, and 23-24. Both represent a straight-forward application of the nonrelativistic CC theory for many-electron systems with Dirac operator replacing the traditional one-electron Schrödinger Hamiltonian, and with the possibility of adding Breit operator to Coulomb interaction.

To summarize, in current relativistic CC theories, the intermediately normalized ground state wave function of the Hamiltonian is related to the DF ground state configuration by an exponential operator containing the cluster operator $\hat{T}$,

$$|\Phi^0_N\rangle = e^{\hat{T}} |\Psi^0_N\rangle \tag{23}$$

such that

$$\langle \Psi^0_N | \Phi^0_N \rangle = \langle \Psi^0_N | e^{\hat{T}} | \Psi^0_N \rangle = 1. \tag{24}$$

Because $|\Phi^0_N\rangle$ is assumed to be the true ground state wave function,

$$(\hat{H}_{D,ext} + \Lambda_+ \hat{H}_C \Lambda_+ - E^0_N)|\Phi^0_N\rangle = E_{correl} |\Phi^0_N\rangle. \tag{25}$$

Moreover, the exponential operator may be expanded in a series containing powers of $\hat{T}$ that leads to a coupling among the terms in the cluster. A large number of terms in the second and higher orders make a nonzero contribution. Using

$$\hat{T} = \hat{T}_2 = \sum_{\substack{m<n \\ r<s}} C^{rs}_{mn} a^\dagger_r a^\dagger_s a_n a_m \tag{26}$$

that is a linear combination of the double excitations, one may write (23) as

$$|\Phi^0_N\rangle = (1 + \hat{T}_2 + \hat{T}_4 + ...)|\Psi^0_N\rangle \tag{27}$$

where

$$\hat{T}_4 = \sum_{\substack{m<m'<n<n' \\ r<s<t<u}} C^{rstu}_{mm'nn'} a_r^\dagger a_s^\dagger a_t^\dagger a_u^\dagger a_{n'} a_n a_{m'} a_m, \tag{28}$$

$$C^{rstu}_{mm'nn'} = C^{rs}_{mm'} C^{tu}_{nn'} - \langle C^{rs}_{mm'} * C^{tu}_{nn'} \rangle,$$

etc. The doubles cluster $\hat{T}_2$ makes the most prominent contribution to the correlation energy and the correlated ground state,

$$\sum_{\substack{m<n \\ r<s}} \langle \Psi_N^0 | \hat{H}_{D,ext} + \Lambda_+ \hat{H}_C \Lambda_+ |^{rs}_{mn} \rangle C^{rs}_{mn} = E_{correl}, \tag{29}$$

while in principle $\hat{T}_4$ improves the calculation by supplying quadruples and determines a better set of coefficients for the doubles from the relation

$$\langle^{rs}_{mn}| \hat{H}_{D,ext} + \Lambda_+ \hat{H}_C \Lambda_+ | \Psi_N^0 \rangle + \sum_{\substack{p<q \\ t<u}} \langle^{rs}_{mn}| \hat{H}_{D,ext} + \Lambda_+ \hat{H}_C \Lambda_+ - E_N^0 |^{tu}_{pq} \rangle C^{tu}_{pq}$$
$$- \sum_{\substack{p<q \\ t<u}} \langle \Psi_N^0 | \hat{H}_{D,ext} + \Lambda_+ \hat{H}_C \Lambda_+ |^{tu}_{pq} \rangle \langle C^{rs}_{mn} * C^{tu}_{pq} \rangle = 0 \tag{30}$$

where the second part of equation (28) and the equation (29) have been used. These are the CC equations involving the ground state wave function in (27).

The choice $\hat{T} = \hat{T}_2$ may be modified by adding a cluster of single excitations

$$\hat{T}_1 = \sum_{m,r} C_m^r a_r^\dagger a_m \tag{31}$$

to the argument of the exponential operator. This gives $\hat{T} = \hat{T}_1 + \hat{T}_2$ and is known in literature as the CCSD. It also offers the CCSDTQ with selective slices of triples and quadruples:

$$| \Phi_N^0 \rangle = (1 + \hat{T}_1 + \hat{T}_2 + \hat{T}_3 + \hat{T}_4 + ...) | \Psi_N^0 \rangle, \tag{32}$$

$$\hat{T}_3 = \sum_{\substack{m<m'<n \\ r<s<t}} C^{rst}_{mm'n} a_r^\dagger a_s^\dagger a_t^\dagger a_n a_{m'} a_m,$$

etc. The series expansion of the exponential operator forces the coefficients in the cluster of a given order, (say, the third or fourth order), to be determined by the lower order coefficients.

## 4. QED Based Coupled Cluster

When one starts from the Hamiltonian operator of QED, one needs to consider a product state vector

$$| \Psi_0 \rangle = | \Psi_N^0 \rangle | \Psi_{rad}^0 \rangle \tag{33}$$

and at least two clusters for a CC treatment, one for the radiative effect and the other for the matter correlation. The following relations are observed:

$$E_N^0 = \langle \Psi_0 | \hat{H}_D^0 + \Lambda_+ \hat{H}_C \Lambda_+ | \Psi_0 \rangle, \tag{34}$$

$$E_{rad}^0 = \langle \Psi_0 | \hat{H}_{rad}^0 | \Psi_0 \rangle. \tag{35}$$

The interaction operator $\hat{H}_{int}^{(2)}$ linearly varies with the photon creation and destruction operators. Its expectation value over the photon ground state $| \Psi_{rad}^0 \rangle$ vanishes. It is bilinear in matter field, and because of the presence of Dirac $\alpha$ matrix operator, it accommodates single-particle excitations. The radiative cluster $\hat{T}_{1,int}^{(2)}$ is to work with $\hat{H}_{int}^{(2)}$ at least as a linear factor and the product should give nonzero average values over the states of radiation as well as matter states. This condition requires the simplest $\hat{T}_{1,\text{int}}^{(2)}$ to be formed from single-particle excitations and to linearly vary with photon operators. Thus it needs to differ from $\hat{H}_{\text{int}}^{(2)}$ only by a multiplicative factor for each intermediate state.

A simple way of guessing the cluster operators is to follow Rayleigh-Schrödinger perturbation theory that is known to be size-extensive order by order. For instance, the second order energy correction $\langle VQV \rangle$ indicates the cluster operator to be of the form $QV$ where $\hat{Q} = \sum_I |I\rangle\langle I|(E_{HF} - H_{HF})^{-1}$, the summation is over the intermediate states, and $V = (\hat{H}_C - \hat{V}_{HF}) = \hat{H}_C - \sum_{i=1}^{N} v_{DF}(r_i)$ in the nonrelativistic theory. A cluster of this type differs from a linear combination where coefficients are determined from a set of matrix equations as in the nonrelativistic CC. The difference is partly caused because of the finite nature of basis set, (that is, one works with eigenvectors and matrix eigenvalues), and mainly arises from the higher order energy terms, (the third order energy correction being $\langle V\hat{Q}V\hat{Q}V \rangle - \langle V \rangle \langle V\hat{Q}\hat{Q}V \rangle$ instead of $\langle V\hat{Q}V\hat{Q}V \rangle/2$, etc.). When the higher order terms are negligibly small and the calculation relies on a complete basis set, the cluster operator would be written as $\hat{T} = \hat{Q}V$. The operator $\hat{H}_{\text{int}}^{(2)}\hat{T}_{1,\text{int}}^{(2)}$ $(= V\hat{Q}V)$ is manifestly hermitean.

Therefore, the radiative cluster is written as

$$\hat{T}_{1,\text{int}}^{(2)} = -\frac{e}{c\sqrt{\Omega}} \sum_n \sum_{k\lambda} \sum_{N'_{k\lambda}} |\Psi_N^n\rangle |N'_{k\lambda}\rangle [E_{DF}^0 - E_{DF}^n + (N_{k\lambda} - N'_{k\lambda})\hbar ck]^{-1} \langle N'_{k\lambda}| \langle \Psi_N^n| \times$$
$$\int d^3r \, \hat{J}_D(r) \cdot \left[ A_{k\lambda}\lambda e^{i(k\cdot r - \omega_k t)} + A_{k\lambda}^\dagger \lambda^* e^{-i(k\cdot r - \omega_k t)} \right]. \tag{36}$$

The operator $\hat{T}_{1,\text{int}}^{(2)}$ is of order $\alpha Z$ while the interaction $\hat{H}_{\text{int}}^{(2)}$ has the order $mc^2\alpha^3 Z^3$. The moments $\langle \hat{T}_{1,\text{int}}^{(2)\,n_1} \rangle_{\Psi_N^0}$ and $\langle \hat{T}_{1,\text{int}}^{(2)\,n_2} \rangle_{\Psi_{rad}^0}$ are both nonzero only when $n_1$ and $n_2$ are even positive numbers including zero.

The double excitation operator $\hat{T}_2$ is a staple for the matter cluster in both nonrelativistic and relativistic CC treatments. It can be fortified by adding the singles cluster $\hat{T}_1$ for a better calculation of the correlation energy. Thus one may choose $\hat{T}_{mat} = \hat{T}_1 + \hat{T}_2$ or simply $\hat{T}_{mat} = \hat{T}_2$. The clusters $\hat{T}_1$, $\hat{T}_2$, $\hat{T}_3$ and $\hat{T}_4$ are neither hermitean nor anti-hermitean. The net cluster in the exponential is written as the sum $(\hat{T}_{1,\text{int}}^{(2)} + \hat{T}_{mat})$.

The first intermediately normalized state is
$$|\tilde{\Phi}_0\rangle = e^{\hat{T}_{mat}} |\Psi_0\rangle = (1 + \hat{T}_1 + \hat{T}_2 + \hat{T}_3 + \hat{T}_4 + ...)|\Psi_0\rangle \tag{37}$$

where $\hat{T}_1$, $\hat{T}_2$, $\hat{T}_3$, $\hat{T}_4$, etc., are given in equations (26)-(32). As the matter clusters consist of excitations from the Dirac-Fock ground state configuration, $\langle \Psi_0 | \tilde{\Phi}_0 \rangle = 1$ that is a variant of equation (24). The final intermediately normalized state can be written as

$$|\Phi_0\rangle = \frac{e^{\hat{T}_{1,int}^{(2)}}}{\langle e^{\hat{T}_{1,int}^{(2)}} \rangle_{\Psi_{rad}^0}} |\tilde{\Phi}_0\rangle = \frac{e^{\hat{T}_{1,int}^{(2)}}}{\langle e^{\hat{T}_{1,int}^{(2)}} \rangle_{\Psi_{rad}^0}} e^{\hat{T}_{mat}} |\Psi_0\rangle \tag{38}$$

such that $\langle \Psi_0 | \Phi_0 \rangle = 1$.

## 5. Effects of Radiative Cluster

A few observations can be made at this juncture. These are as follows:
(1) When the averaging is done over the reference state of photons,

$$\langle \hat{H}_{QED} e^{\hat{T}_{1,int}^{(2)}} \rangle_{\Psi_{rad}^0} = [\hat{H}_D^0 + \hat{H}_C + E_{rad}^0]\langle e^{\hat{T}_{1,int}^{(2)}} \rangle_{\Psi_{rad}^0} + \\
+ \langle \Psi_{rad}^0 | H_{int}^{(2)} \hat{T}_{1,int}^{(2)} (1 + \frac{1}{3!}\hat{T}_{1,int}^{(2)2} + ...) | \Psi_{rad}^0 \rangle. \tag{39}$$

(2) Also,

$$\langle e^{\hat{T}_{1,int}^{(2)}} \rangle_{\Psi_{rad}^0} = 1 + \frac{1}{2!}\langle \hat{T}_{1,int}^{(2)2} \rangle_{\Psi_{rad}^0} + O(\alpha^4 Z^4), \tag{40}$$

so that

$$\frac{\langle \hat{H}_{QED} e^{\hat{T}_{1,int}^{(2)}} \rangle_{\Psi_{rad}^0}}{\langle e^{\hat{T}_{1,int}^{(2)}} \rangle_{\Psi_{rad}^0}} = E_{rad}^0 + H_D^0 + \hat{H}_C + [\langle \Psi_{rad}^0 | H_{int}^{(2)} \hat{T}_{1,int}^{(2)} | \Psi_{rad}^0 \rangle + O(mc^2\alpha^6 Z^6)]. \tag{41}$$

When one puts all the pieces together, one obtains the effect of the interaction of two electrons by absorbing (emitting) and subsequently emitting (absorbing) a transverse virtual photon,

$$\langle \Psi_{rad}^0 | \hat{H}_{int}^{(2)} \hat{T}_{1,int}^{(2)} | \Psi_{rad}^0 \rangle = \frac{e^2}{c^2\Omega} \int d^3r \int d^3r' \, \hat{J}_D(r) \cdot \\
\sum_{k\lambda} \sum_{N'_{k\lambda}} \langle N_{k\lambda} | \left[ A_{k\lambda}\lambda e^{i(k\cdot r - \omega_k t)} + A_{k\lambda}^\dagger \lambda^* e^{-i(k\cdot r - \omega_k t)} \right] | N'_{k\lambda} \rangle \times \\
\sum_n |\Psi_N^n\rangle [E_{DF}^0 - E_{DF}^n + (N_{k\lambda} - N'_{k\lambda})\hbar c k]^{-1} \langle \Psi_N^n | \hat{J}_D(r') \cdot \\
\langle N'_{k\lambda} | \left[ A_{k\lambda}\lambda e^{i(k\cdot r' - \omega_k t)} + A_{k\lambda}^\dagger \lambda^* e^{-i(k\cdot r' - \omega_k t)} \right] | N_{k\lambda} \rangle. \tag{42}$$

The photon matrix elements can be easily calculated by using equations (14) and (15). The sum over the discrete variable $k$ for a finite $\Omega$ can be replaced by an integral over the continuous variable $k$ in the limit of infinite volume as shown below,

$$\Omega^{-1} \sum_k \to \frac{1}{8\pi^3} \int d^3k. \tag{43}$$

Furthermore, when the polarization vectors are considered in real forms, one obtains

$$\langle \Psi_{rad}^0 | H_{int}^{(2)} \hat{T}_{1,int}^{(2)} | \Psi_{rad}^0 \rangle$$

$$= \frac{e^2 \hbar}{4\pi^2 c} \int \frac{d^3k}{k} \sum_\lambda \Bigl[ \int d^3r \int d^3r' e^{i\mathbf{k}\cdot(\mathbf{r}-\mathbf{r}')} \times$$

$$\sum_n \hat{J}_D(\mathbf{r})\cdot\lambda | \Psi_N^n \rangle \frac{N_{k\lambda}+1}{E_{DF}^0 - E_{DF}^n - \hbar ck} \langle \Psi_N^n | \hat{J}_D(\mathbf{r}')\cdot\lambda \qquad (44)$$

$$+ \int d^3r \int d^3r' e^{-i\mathbf{k}\cdot(\mathbf{r}-\mathbf{r}')} \times$$

$$\sum_n \hat{J}_D(\mathbf{r})\cdot\lambda | \Psi_N^n \rangle \frac{N_{k\lambda}}{E_{DF}^0 - E_{DF}^n + \hbar ck} \langle \Psi_N^n | \hat{J}_D(\mathbf{r}')\cdot\lambda \Bigr].$$

Further simplification can be achieved by discarding most of the electron self-energy in presence of the external potential that represents interaction with the positively charged nuclei, while keeping only the part that exists even in radiation vacuum ($\hat{H}_{SE,vac}^{ext}$), and considering that for the remaining term the virtual photon energy is much greater than the excitation energy, $\hbar ck \gg | E_{DF}^n - E_{DF}^0 |$, so that the denominators can be approximated and the sum over $N$-electron states can be replaced by unity leading to the operator $\hat{H}_{Breit}$. Thus, after correcting for the electron self-energy,

$$\langle \Psi_{rad}^0 | \hat{H}_{int}^{(2)} \hat{T}_{1,int}^{(2)} | \Psi_{rad}^0 \rangle \doteq \hat{H}_{SE,vac}^{ext} + \hat{H}_{Breit} \qquad (45)$$

where

$$\hat{H}_{SE,vac}^{ext} = \frac{e^2 \hbar}{4\pi^2 c} \int \frac{d^3k}{k} \sum_\lambda \int d^3r \sum_n \hat{J}_D(\mathbf{r})\cdot\lambda \frac{|\Psi_N^n\rangle\langle\Psi_N^n|}{E_{DF}^0 - E_{DF}^n - \hbar ck} \hat{J}_D(\mathbf{r})\cdot\lambda, \qquad (46)$$

$$\hat{H}_{Breit} = -\frac{e^2}{4\pi^2 c^2} \int \frac{d^3k}{k^2} \sum_\lambda \Bigl[ (N_{k\lambda}+1)\int d^3r \int d^3r' e^{i\mathbf{k}\cdot(\mathbf{r}-\mathbf{r}')} \hat{J}_D(\mathbf{r})\cdot\lambda \hat{J}_D(\mathbf{r}')\cdot\lambda$$
$$- N_{k\lambda} \int d^3r \int d^3r' e^{-i\mathbf{k}\cdot(\mathbf{r}-\mathbf{r}')} \hat{J}_D(\mathbf{r})\cdot\lambda \hat{J}_D(\mathbf{r}')\cdot\lambda \Bigr]. \qquad (47)$$

Lamb Shift

The energy difference $| E_{DF}^n - E_{DF}^0 |$ may be kept in the denominator in $\hat{T}_{1,int}^{(2)}$ while retaining the one-particle self-energy accompanied by scattering, that is, one scattering of the electron by the external potential preceded by the emission (absorption) of a virtual photon and followed by the absorption (emission) of the same photon. The energy contribution that evolves from $\hat{H}_{SE,vac}^{ext}$ is already of order less than $\alpha Z$, and it eventually leads to the renormalization of mass and gives rise to the effect historically known as Lamb shift.

Averaging over the transverse polarization yields a factor of ⅔ in equation (46) and the self-energy contribution can be written as

$$\hat{H}_{SE,vac}^{ext} = -\frac{2\alpha\hbar}{3\pi c} \int_0^\infty dk \int d^3r \Bigl[ \hat{J}_D(\mathbf{r})\cdot \hat{J}_D(\mathbf{r})$$
$$- \sum_n \hat{J}_D(\mathbf{r}) | \Psi_N^n \rangle \cdot \langle \Psi_N^n | \hat{J}_D(\mathbf{r}) \frac{E_{DF}^n - E_{DF}^0}{E_{DF}^n - E_{DF}^0 + \hbar ck} \Bigr]. \qquad (48)$$

This integral diverges. The fundamental point is that when (48) is translated into the nonrelativistic limit, a renormalization of mass is seen necessary and the same task is achieved by subtracting a similar correction for the free electron. The latter correction is given by the integral in (48) where only the $n = 0$ term is retained in the sum within the integrand. Thus the visible part of this self-energy is given by the difference

$$\Delta \hat{H}_{SE,vac} = \hat{H}_{SE,vac}^{ext} - \hat{H}_{SE,vac}^{free}$$

$$= \frac{2\alpha\hbar}{3\pi c} \int_0^\infty dk \int d^3r \sum_{n(\neq 0)} \hat{J}_D(r) |\Psi_N^n\rangle \cdot \langle \Psi_N^n | \hat{J}_D(r) \frac{E_{DF}^n - E_{DF}^0}{E_{DF}^n - E_{DF}^0 + \hbar ck} \quad (49)$$

which is only logarithmically divergent. The logarithmic divergence is removed by the intelligent use of a cut-off $k_{co}=mc/\hbar$ as the upper boundary of the $k$-integral such that for a real-life calculation, one is left with an effective Hamiltonian operator for Lamb shift,

$$\hat{H}_{Lamb} = \frac{2\alpha}{3\pi c^2} \int d^3r \sum_{n'(\neq 0)} \hat{J}_D(r) |\Psi_N^{n'}\rangle \cdot \langle \Psi_N^{n'} | \hat{J}_D(r) \times$$

$$(E_{DF}^{n'} - E_{DF}^0) \ln\left(\frac{E_{DF}^{n'} - E_{DF}^0 + \hbar ck_{CO}}{E_{DF}^n - E_{DF}^0}\right) \quad (50)$$

A general expression valid for an arbitrary DF state, (say, for the $n$th state), can be written as

$$\hat{H}_{Lamb} = \frac{2\alpha}{3\pi c^2} \int d^3r \sum_{n'(\neq n)} \hat{J}_D(r) |\Psi_N^{n'}\rangle \cdot \langle \Psi_N^{n'} | \hat{J}_D(r) \times$$

$$(E_{DF}^{n'} - E_{DF}^n) \ln\left(\frac{mc^2}{|E_{DF}^{n'} - E_{DF}^n|}\right) \quad (51)$$

where use has been made of $\hbar ck_{co} = mc^2 >> |E_{DF}^{n'} - E_{DF}^n|$. The $k$-integral is in reality a principal value integral, and this leads to the absolute value in (51). This expression is easily amenable to the calculation of Lamb shift as average over the Dirac-Fock $n$th bound state configuration. As a classic example (albeit in the one-electron case), the $2S_{1/2}$–$2P_{1/2}$ shift in hydrogen atom is about 1057 MHz.

Breit interaction

For any specific $k$ vector, the two space integrals over the exponential functions in equation (47) are equal by symmetry. A sum over the polarization vector is carried out to get

$$\hat{H}_{Breit} = -\frac{e^2}{4\pi^2 c^2} \int \frac{d^3k}{k^2} \int d^3r \int d^3r' \, e^{i\mathbf{k}\cdot(\mathbf{r}-\mathbf{r}')} \left[\hat{J}_D(r)\cdot\hat{J}_D(r') - \frac{\hat{J}_D(r)\cdot\mathbf{k}\,\hat{J}_D(r')\cdot\mathbf{k}}{k^2}\right] \quad (52)$$

that is a sum of magnetic and retarded interactions corresponding to the two terms in the square bracket in the integrand. Singer transformation gives the first part of the integral as

$$\hat{H}_{Magnetic} = -\frac{e^2}{2c^2} \int d^3r \int d^3r' \frac{\hat{J}_D(r)\cdot\hat{J}_D(r')}{|\mathbf{r}-\mathbf{r}'|} \quad (53)$$

while the equality

$$\frac{1}{2\pi^2} \int \frac{d^3k}{k^2} e^{i\mathbf{k}\cdot\mathbf{r}} \frac{\mathbf{a}\cdot\mathbf{k}\,\mathbf{b}\cdot\mathbf{k}}{k^2} = \frac{1}{2r}\left[\mathbf{a}\cdot\mathbf{b} - \frac{\mathbf{a}\cdot\mathbf{r}\,\mathbf{b}\cdot\mathbf{r}}{r^2}\right] \quad (54)$$

reduces the second part of interaction into the form

$$\hat{H}_{Retarded} = \frac{e^2}{4c^2} \int d^3r \int d^3r' \frac{1}{|\mathbf{r}-\mathbf{r}'|}\left[\hat{J}_D(r)\cdot\hat{J}_D(r') - \frac{\hat{J}_D(r)\cdot(\mathbf{r}-\mathbf{r}')\,\hat{J}_D(r')\cdot(\mathbf{r}-\mathbf{r}')}{|\mathbf{r}-\mathbf{r}'|^2}\right]. \quad (55)$$

Therefore, Breit interaction can be written as the sum

$$\hat{H}_{Breit} = -\frac{e^2}{4c^2} \int d^3r \int d^3r' \frac{1}{|\boldsymbol{r}-\boldsymbol{r}'|} \left[ \hat{\boldsymbol{J}}_D(\boldsymbol{r}) \cdot \hat{\boldsymbol{J}}_D(\boldsymbol{r}') + \frac{\hat{\boldsymbol{J}}_D(\boldsymbol{r}) \cdot (\boldsymbol{r}-\boldsymbol{r}') \hat{\boldsymbol{J}}_D(\boldsymbol{r}') \cdot (\boldsymbol{r}-\boldsymbol{r}')}{|\boldsymbol{r}-\boldsymbol{r}'|^2} \right]. \quad (56)$$

After self-energy corrections, it reduces to its usual form

$$\hat{H}_{Breit} = \sum_{1 \leq i < j \leq N} B(i,j), \quad (57)$$

$$B(i,j) = -\frac{e^2}{2r_{ij}} \left[ \boldsymbol{\alpha}_i \cdot \boldsymbol{\alpha}_j + \frac{\boldsymbol{\alpha}_i \cdot \boldsymbol{r}_{ij} \boldsymbol{\alpha}_j \cdot \boldsymbol{r}_{ij}}{r_{ij}^2} \right] \quad (58)$$

in the notation of "first" quantization. The Breit interaction energy in the ground state of a light atom such as helium is of the order of $10^5$ MHz, and for neon it is about $10^8$ MHz. To compare, the $2p_{1/2}$-$2p_{3/2}$ fine structure in hydrogen atom is $1.095 \times 10^4$ MHz.

The hyperfine interaction

The hyperfine correction is another QED effect, an additional magnetic interaction to be accommodated within the Dirac-Coulomb-Breit Hamiltonian. Following the two-fermion formulation of Chraplyvy[26-27] and a subsequent development by Barker and Glover,[28] this interaction between the electrons and the fermion nuclei in a molecule is written as

$$\hat{H}_{hf} = -\sum_{\substack{i=1 \\ (electron)}}^{N} \sum_{\substack{n \\ (nucleus)}}^{fermion} \beta_i \beta_n \frac{Z_n e^2 \hbar^2 g_e g_n}{4mM_n c^2} \times$$

$$\left( \frac{\boldsymbol{\sigma}_{Di} \cdot \boldsymbol{\sigma}_{Dn}}{|\boldsymbol{r}_i - \boldsymbol{R}_n|^3} - 3 \frac{\boldsymbol{\sigma}_{Di} \cdot (\boldsymbol{r}_i - \boldsymbol{R}_n) \boldsymbol{\sigma}_{Dn} \cdot (\boldsymbol{r}_i - \boldsymbol{R}_n)}{|\boldsymbol{r}_i - \boldsymbol{R}_n|^5} - \frac{8\pi}{3} \boldsymbol{\sigma}_{Di} \cdot \boldsymbol{\sigma}_{Dn} \delta^3(\boldsymbol{r}_i - \boldsymbol{R}_n) \right). \quad (59)$$

In the above $\sigma_D$ stands for the Dirac spin matrix vector. The first two terms within the bracket in (62) represent the dipolar interaction between the electron spin and the nuclear spin at a finite distance. The third term is known as the Fermi contact interaction. The hyperfine splitting is of order $m^2c^2\alpha^4Z^{1-3}/M \sim m\alpha^2 Z^{1-3}/M$ hartree. To give an estimate of the order of magnitude, the hyperfine splitting of the hydrogen atom is 1420 MHz in its ground state, 177 MHz in $2S_{1/2}$ state and 59 MHz in $2p_{1/2}$ and $2p_{3/2}$ states. Hyperfine structure of $Cd^+$ has been calculated by Li et al. using the relativistic CC.[29]

These results allow the definition of an effective Hamiltonian of QED from equation (41)

$$\hat{H}_{QED}^{eff} = \langle \hat{H}_{QED} e^{\hat{T}_{1,int}^{(2)}} \rangle_{\Psi_{rad}^0} / \langle e^{\hat{T}_{1,int}^{(2)}} \rangle_{\Psi_{rad}^0}$$

$$= \left( \hat{H}_D^0 + \Lambda_+ \hat{H}_C \Lambda_+ + \hat{H}_C^{Pair} + E_{rad}^0 + \Lambda_+ \hat{H}_{Breit} \Lambda_+ + \hat{H}_{Lamb} + \hat{H}_{hf} \right) \quad (60)$$

where the hyperfine interaction has been added to complement the electronic Breit operator,[28,33] and following Sucher's suggestion, Breit operator has been considered in the projected form.

Additional QED correction terms are known to arise from the polarization of vacuum due to the creation and annihilation of virtual electron-positron pairs using the operator $\hat{H}_C^{Pair}$. Such corrections are mostly blocked on the ground of the exclusion principle. After the Pauli blocking, the 1-pair and 2-pair contributions to energy appear as tiny positive corrections of orders $mc^2\alpha^6 Z^6$ and $mc^2\alpha^8 Z^8$, respectively. The pair terms do not appear in a relativistic configuration interaction (RCI) calculation that is based on the configurations prepared from only the DF positive-energy eigenvectors (PERCI). The 1-pair (and 2-pair) term(s) appear(s) when the all-energy eigenvectors are considered (AERCI), that is, the spurious solutions of negative energy from the DF calculation are included to obtain de-excitations from the ground state configuration in the RCI. The AERCI corresponds to a many-electron min-max procedure, and the (AERCI – PERCI) energy difference was found to be in excellent agreement with an analytical estimate of the 1-pair energy [30]. The vacuum polarization effect on energy is fundamentally a correlation effect, and it can be realized from the cluster operator technique if one considers the more complete Coulomb interaction while exploring the influence of a more detailed matter cluster.

## 6. Matter Clusters

Just as arbitrary free particle 4-component wave functions need both positive-energy and negative-energy eigenspinors for completeness, the one-electron bound state solutions in Furry or Dirac-Fock picture require not only the positive-energy eigenvectors but also the spurious solutions to form an orthonormal complete set. Thus any arbitrary trial spinor must be written as a linear combination of the eigenvectors of both positive and negative energy. This leads to the possibility of variation collapse[31-32] and to avoid the same problem one may resort to a min-max principle for solving the involved wave equation.[33-35] It has been assumed here that the Dirac-Fock orbitals have already been obtained from a min-max principle such as the one discussed in ref. 33.

It is normally taken for granted that the positive-energy Dirac eigenspinors representing bound states, (that is, Dirac or DF eigenvectors of positive energy), form a complete space for the bound state solutions so that the one-electron projector $\lambda_+$ can be used to form the $N$-electron projection operator $\Lambda_+$ that in turn is used to build the projected interaction. This assumption is of course wrong though it has been deeply entrenched in quantum chemical calculations. The reason for the wrong assumption is that Sucher preferred to work in the free particle picture where positive and negative energy solutions are distinctly known and the completeness relations hold separately for them. It was Sucher himself who showed that the non-perturbative use of the interaction associated with the Feynman gauge photon propagator in place of the interaction associated with the Coulomb gauge propagator leads to energy levels that are incorrect at the level of atomic fine structure.[36] Indeed the projector $\lambda_-(i) = \underset{v'}{S} |v'(i)\rangle\langle v'(i)|$ where $v'$ stands for the spurious solutions of negative energy, contributes to an arbitrary trial spinor in the positive energy range at order $(p/mc) \sim \alpha Z$ so that the energy levels become incorrect at order $mc^2 \alpha^4 Z^4$. Liu and Lindgren have discussed quantum chemistry beyond the no-pair Hamiltonian,[37] and calculations on superheavy elements including the QED effects beyond the no-pair Hamiltonian have been reported by Schwerdtfeger et al.[38]

In a finite basis calculation, it would be proper to include the spurious yet square integrable eigenvectors to form an approximation to the pair operator:

$$\hat{H}_C^{Pair} = \Lambda \hat{H}_C \Lambda - \Lambda_+ \hat{H}_C \Lambda_+,$$

$$\Lambda = \prod_{i=1}^{N} \lambda(i), \quad \lambda(i) = \lambda_+(i) + \lambda_-(i). \tag{61}$$

Correlation effects are determined from the matter clusters. Equations (22) and (35) together give

$$\langle\Psi_0|\hat{H}_{QED}|\Psi_0\rangle=\langle\Psi_N^0|\hat{H}_{rad}^0+\hat{H}_{D,ext}+\Lambda_+\hat{H}_C\Lambda_++\hat{H}_C^{Pair}+\hat{H}_{int}^{(2)}|\Psi_N^0\rangle=E_{rad}^0+E_N^0 \quad (62)$$

whereas (60) yields

$$\langle\Psi_N^0|\hat{H}_{QED}^{eff}|\Psi_N^0\rangle=(E_N^0+E_{rad}^0)+(E_{Breit}^0+E_{Lamb}^0+E_{Hf}^0), \quad (63)$$

$E_{Breit}^0$, $E_{Lamb}^0$ and $E_{Hf}^0$ being the expectation values of $\hat{H}_{Breit}$, $\hat{H}_{Lamb}$ and $\hat{H}_{Hf}$ respectively over the Dirac-Fock ground state configuration $\Psi_N^0$. Combining equations (38), (41), (60) and (63), one obtains

$$\langle\Psi_{rad}^0|\hat{H}_{QED}-[(E_N^0+E_{rad}^0)+\langle(\hat{H}_{Breit}+\hat{H}_{Lamb}+\hat{H}_{hf})\rangle]|\Phi_0\rangle$$
$$=\left(\hat{H}_{QED}^{eff}-[(E_N^0+E_{rad}^0)+(E_{Breit}^0+E_{Lamb}^0+E_{Hf}^0)]\right)e^{\hat{T}_{mat}}|\Psi_N^0\rangle \quad (64)$$
$$=E_{correl}e^{\hat{T}_{mat}}|\Psi_N^0\rangle$$

where $\hat{T}_{mat}$ is extended to cater for the pair operator in (61). A hierarchy of matrix equations can be derived from (64) and then solved.

The procedure can be illustrated by using only the doubles in the matter cluster. However, the cluster now includes not only the excitations from the DF ground state configuration to the conventional virtual orbitals but also a mixture of excitations to the virtuals and de-excitations to the spurious levels (indicated by primes) and even double deexcitations,

$$\hat{T}_{mat}=\hat{T}_2+\hat{T}_2^{1-pair}+\hat{T}_2^{2-pair},$$
$$\hat{T}_2=\sum_{\substack{m<n\\r<s}}C_{mn}^{rs}a_r^\dagger a_s^\dagger a_n a_m, \quad (65)$$
$$\hat{T}_2^{1-pair}=\sum_{\substack{m<n\\r,s'}}C_{mn}^{r\,s'}a_r^\dagger a_{s'}^\dagger a_n a_m, \quad \hat{T}_2^{2-pair}=\sum_{\substack{m<n\\r'<s'}}C_{mn}^{r'\,s'}a_{r'}^\dagger a_{s'}^\dagger a_n a_m.$$

The 1-pair and 2-pair clusters are evident. Similar de-excitations were included in AERCI [25]. Equation (29) is translated in the present treatment as

$$\sum_{\substack{m<n\\r<s}}\langle\Psi_N^0|\Lambda_+\hat{H}_C\Lambda_++\hat{H}_{Breit}+\hat{H}_{Lamb}+\hat{H}_{Hf}|_{mn}^{rs}\rangle C_{mn}^{rs}=E_{correl}, \quad (66)$$

$$\sum_{\substack{m<n\\r,\,s'}}\langle\Psi_N^0|\hat{H}_C^{Pair}|_{mn}^{r\,s'}\rangle C_{mn}^{r\,s'}=\delta E_{1-pair}, \quad (67)$$

$$\sum_{\substack{m<n\\r',\,s'}}\langle\Psi_N^0|\hat{H}_C^{Pair}|_{mn}^{r'\,s'}\rangle C_{mn}^{r'\,s'}=\delta E_{2-pair}, \quad (68)$$

whereas the equation corresponding to (30) appears as

$$\langle_{mn}^{rs}|\hat{H}_D^0+\Lambda_+\hat{H}_C\Lambda_++\hat{H}_{Breit}+\hat{H}_{Lamb}|\Psi_N^0\rangle$$
$$+\sum_{\substack{p<q\\t<u}}\langle_{mn}^{rs}|\hat{H}_D^0+\Lambda_+\hat{H}_C\Lambda_++\hat{H}_{Breit}+\hat{H}_{Lamb}+\hat{H}_{hf}-(E_N^0+E_{Breit}^0+E_{Lamb}^0+E_{Hf}^0)|_{pq}^{tu}\rangle C_{pq}^{tu} \quad (69)$$
$$-\sum_{\substack{p<q\\t<u}}\langle\Psi_N^0|\hat{H}_D^0+\Lambda_+\hat{H}_C\Lambda_++\hat{H}_{Breit}+\hat{H}_{Lamb}|_{pq}^{tu}\rangle\langle C_{mn}^{rs}*C_{pq}^{tu}\rangle=0.$$

These equations in (69) can be solved to obtain the coefficients and then obtain the correlation energy from (66). The pair energies can be determined by using the MBPT expressions for the involved coefficients. Very good estimates are obtained from the approximations $C_{mn}^{r\,s'} = \langle rs' \| mn \rangle / 2mc^2$ and $C_{mn}^{r'\,s'} = \langle r's' \| mn \rangle / 4mc^2$. It is easy to find

$$\delta E_{1-pair} = \frac{1}{2mc^2} \sum_{\substack{m<n \\ r,\,s'}} |\langle rs' \| mn \rangle|^2 \sim O(mc^2 \alpha^6 Z^6),$$

$$\delta E_{2-pair} = \frac{1}{4mc^2} \sum_{\substack{m<n \\ r',\,s'}} |\langle r's' \| mn \rangle|^2 \sim O(mc^2 \alpha^8 Z^8). \tag{70}$$

While Breit interaction was added to QFT-based CC as an afterthought, here it is directly involved. Its action is at par with that of Coulomb interaction, though smaller in the absolute magnitude by an order of $\alpha^2 Z^2$. Hence the coefficients and the exponential cluster $\hat{T}_2$ are determined by it. The pair clusters do not affect Breit interaction as the latter was obtained for negligibly small energy differences in the denominator, $(\hbar ck)^{-1} | E_{DF}^n - E_{DF}^0 | \ll 1$. The Lamb shift is also updated as a natural recourse to the many-body level. Nonzero matrix elements of the charge current can be obtained from two states differing by a single excitation. Hence singly excited intermediate electronic states in $\hat{H}_{Lamb}$ can contribute to both (63) and (64). The electron-nucleus hyperfine interactions are one-electron effects and they can contribute to the correlation energy and the correlated wave function through the second term in (69), thereby modifying the coefficients and subsequently updating the correlation energy in (66).

These contributions would be of course more extensive in CCSD and its derivative procedures that include some of the higher order excitations.

7. **Conclusions**

For lighter atoms, intricate spectroscopic features (such as energy ordering of electronic states and the spin-orbit splitting), and additional radiative effects (such as level shifts due to the retarded interaction with a virtual photon, the Lamb-Retherford effect and the hyperfine splitting) can be observed and compared with theory. However, as mentioned in the introductory section, the radiative effects can be partly concealed in a molecule because of extensive rotational, vibrational and ro-vibronic contributions to total internal energy.

A few observations can be made now:
(1) The effective cluster considered in the present work has been
$$\hat{T}_{eff} = (\hat{T}_{1,\text{int}}^{(2)} + ...)(1 + \hat{T}_1 + \hat{T}_2 + \hat{T}_3 + \hat{T}_4 + ...) + (\hat{T}_1 + \hat{T}_2 + \hat{T}_3 + \hat{T}_4 + ...) + (\hat{T}_2^{1-pair} + \hat{T}_2^{2-pair}) \tag{71}$$
such that $|\Phi_N^0\rangle = (1 + \hat{T}_{eff})|\Psi_N^0\rangle$.
(2) The tactic employed has been to first calculate an average over the radiation state so that matter, radiative and pair effects become separated:
$$\langle \hat{H}_{QED} \hat{T}_{eff} \rangle_{\Psi_{rad}^0} = [\hat{H}_D^0 + \Lambda_+ \hat{H}_C \Lambda_+ + E_{rad}^0](1 + \hat{T}_1 + \hat{T}_2 + \hat{T}_3 + \hat{T}_4 + ...)$$
$$+ \left(\hat{H}_{Breit} + \hat{H}_{Lamb} + \hat{H}_{hf}\right)(1 + \hat{T}_1 + \hat{T}_2 + \hat{T}_3 + \hat{T}_4 + ...) \tag{72}$$
$$+ \hat{H}_C^{Pair}(\hat{T}_2^{1-pair} + \hat{T}_2^{2-pair}).$$
(3) The factor 1 gives the mean field energy values in addition to the radiation energy.

(4) The matter clusters do not contribute to the radiation energy. Instead, they are responsible for the correlation effects.

(5) Of course it would be possible to accommodate $\hat{T}^{(2)}_{1,\text{int}}(1+\hat{T}_1+\hat{T}'_2)$ in the argument of the exponential of radiative cluster. The interaction $\hat{H}^{(2)}_{\text{int}}$ operates on the exponential operator $e^{\hat{T}^{(2)}_{1,\text{int}}(1+\hat{T}_1+\hat{T}'_2)}$, while the external field Dirac operator and Coulomb interaction together operate on $e^{(\hat{T}_1+\hat{T}'_2)} = 1+\hat{T}_1+\hat{T}_2+\hat{T}_3+\hat{T}_4+...$ The subsequent treatment would be devoid of the simplicity of the present work, though the final result would remain unchanged.

(6) What is new in the present work? One newness is to get the three QED interactions (Lamb, Breit and hyperfine) from a single procedure based on the radiative cluster. Another is to get the pair energy from the matter cluster formalism. As a third novelty, the correlation energy contributions arising from the second line (radiative effects) and the third line (pair terms) of equation (72) appear as additions to the correlation energy in the currently practised relativistic CC.

Consider the example of N' noninteracting minimal-basis $H_2$ molecules with $N = 2N'$ electrons. This system has been treated at the nonrelativistic level (in the limit $c\rightarrow\infty$) in the text by Szabo and Ostlund.[39] A relativistic and electrodynamical version is discussed here. Each molecule has two sets of doubly degenerate Dirac-Fock 4-component spinor orbitals ($\psi_{1\uparrow}$, $\psi_{1\downarrow}$) and ($\psi_{2\uparrow}$, $\psi_{2\downarrow}$) corresponding to the bonding and antibonding sigma molecular orbitals of the nonrelativistic theory. The nature of these spinors is shown, all the associated terms are defined, and integrals are given in Appendix I. The bonding spinors are fully occupied in the DF ground state configuration. Because the molecules do not interact with one another, there are only N' doubles $\left| \begin{array}{c} 2_i \bar{2}_i \\ 1_i \bar{1}_i \end{array} \right\rangle$, ($i = 1,2,\ldots,$ N'), with equal coefficients $C$ for each double in the expanded matter cluster. It is transparent that there is no intermediate state to connect with DF ground state configuration through the 3-current operator and the contribution of $\hat{H}_{Lamb}$ is zero in equations (63) and (66). Also, the hyperfine corrections for two different electron spins cancel each other in the ground state. Therefore, the correlation energy is determined only from Coulomb and Breit interactions:

$$E_{correl} = N'C(K_{12} + K^B_{12}) = N'\left[\Delta - [\Delta^2 + (K_{12} + K^B_{12})^2]^{1/2}\right]. \qquad (73)$$

It is easy to determine the coefficient $C$ from (69).

The example being very familiar from the nonrelativistic theory, what is important here is to get an estimate of relativistic and QED corrections to various energy values and wave functions. These are shown in Appendix I. The nonrelativistic energies (energies in the limit $c\rightarrow\infty$) are shown in equation (I.9). Familiar relativistic corrections such as the kinetic energy correction and the Darwin term are given in equation (I.10). Because the orbital angular momentum is zero in each orbital of the minimal basis calculation, the spin-orbit interaction is absent in this case. The QED corrections to energy values appear only in the form of Breit integrals as shown in equation (I.11). The Lamb corrections do not materialize because of the want of nonzero orbital angular momentum states, while the total of hyperfine interaction energies given in equations (I.12) through (I.14) become zero for a closed shell. Relativistic correction to correlation energy is given in equations (I.15) and (I.16). To order $mc^2\alpha^4Z^4$, the only QED correction to correlation energy appears from Breit interaction as shown in equations (I.17) and (I.18). The lowest order vacuum polarization effects in this example are shown in (I.19).

This work has been strictly limited to the basic theory. Detailed treatments necessary for the open-shell CC (multireference CC) or a state-specific CC are still to be worked out. Also, application has been limited to the simplest exemplary system of the minimal basis hydrogen molecules. As mentioned earlier, methodologies have been established for relativistic extension of CCA, and numerical results have been generated by different workers in this field.[13-20, 23-24, 38] It would be interesting to evaluate the QED contribution to correlation effects and to compare the net QED effects with the molecular energetics at a sufficiently low temperature where the rotational, vibrational and ro-vibronic activities mostly remain frozen.

**Declarations**

The Author declares no conflict of interest.

**Appendix I**

For N' noninteracting minimal basis $H_2$ molecules, the upper and lower components of the 4-component spinors $\psi_{1\uparrow i}$, $\psi_{1\downarrow i}$, $\psi_{2\uparrow i}$ and $\psi_{2\downarrow i}$ (for each molecule numbered as $i$ where $i = 1, \ldots, N'$) are written as

$$u_1 = N_1 \begin{pmatrix} 1\sigma \\ 0 \end{pmatrix}, \quad l_1 = c[\varepsilon_1 + mc^2 - eA_{ext}]^{-1} \boldsymbol{\sigma} \cdot \boldsymbol{p} u_1 \doteq \frac{N_1}{2mc} \begin{pmatrix} p_z(1\sigma) \\ (p_x + ip_y)(1\sigma) \end{pmatrix}, \quad (I.1)$$

$$u_{\bar{1}} = N_1 \begin{pmatrix} 0 \\ 1\sigma \end{pmatrix}, \quad l_{\bar{1}} = c[\varepsilon_1 + mc^2 - eA_{ext}]^{-1} \boldsymbol{\sigma} \cdot \boldsymbol{p} u_{\bar{1}} \doteq \frac{N_1}{2mc} \begin{pmatrix} (p_x - ip_y)(1\sigma) \\ -p_z(1\sigma) \end{pmatrix},$$

$$u_2 = N_2 \begin{pmatrix} 1\sigma^* \\ 0 \end{pmatrix}, \quad l_2 = c[\varepsilon_2 + mc^2 - eA_{ext}]^{-1} \boldsymbol{\sigma} \cdot \boldsymbol{p} u_2 \doteq \frac{N_2}{2mc} \begin{pmatrix} p_z(1\sigma^*) \\ (p_x + ip_y)(1\sigma^*) \end{pmatrix},$$

$$u_{\bar{2}} = N_2 \begin{pmatrix} 0 \\ 1\sigma^* \end{pmatrix}, \quad l_{\bar{2}} = c[\varepsilon_2 + mc^2 - eA_{ext}]^{-1} \boldsymbol{\sigma} \cdot \boldsymbol{p} u_{\bar{2}} \doteq \frac{N_2}{2mc} \begin{pmatrix} (p_x - ip_y)(1\sigma^*) \\ -p_z(1\sigma^*) \end{pmatrix}.$$

where $1\sigma$ and $1\sigma^*$ are nonrelativistic-type orthonormal molecular orbitals of appropriate symmetries, and $\boldsymbol{\sigma}$ is the Pauli spin matrix vector. The normalization constants are $N_{1,2} = (1+\langle p^2 \rangle_{1\sigma,1\sigma^*}/4m^2c^2)^{-1/2} \sim (1-\langle p^2 \rangle_{1\sigma,1\sigma^*}/8m^2c^2)$.

The electronic configuration $\{\psi_{1\uparrow i}\psi_{1\downarrow i}\}$ with $N = 2N'$ gives the mean field energy $E_{MF}^0 = E_N^0 + E_{Breit}^0$ as both $E_{Lamb}^0 = 0$ and $E_{Hf}^0 = 0$. Here

$$E_N^0 = N'E_2^0, \quad E_{Breit}^0 = N'J_{11}^B$$

$$\varepsilon_1 = (h_D)_{11} + (J_{11} + J_{11}^B)$$

$$\varepsilon_2 = (h_D)_{22} + 2(J_{12} + J_{12}^B) - (K_{12} + K_{12}^B)$$

$$E_2^0 = 2(h_D)_{11} + (J_{11} + J_{11}^B) = 2\varepsilon_1 - (J_{11} + J_{11}^B)$$

$$E_2^* = 2(h_D)_{22} + (J_{22} + J_{22}^B) = 2\varepsilon_2 + (J_{22} + J_{22}^B) - 4(J_{12} + J_{12}^B) + 2(K_{12} + K_{12}^B) \quad (I.2)$$

$$J_{11} = <11|11> = K_{11}, \quad J_{12} = <12|12>, \quad K_{12} = <12|21>$$

$$J_{11}^B = <11|\hat{B}(1,2)|11> = K_{11}^B, \quad J_{22}^B = <22|\hat{B}(1,2)|22> = K_{22}^B$$

$$J_{12}^B = <12|\hat{B}(1,2)|12>, \quad K_{12}^B = <12|\hat{B}(1,2)|21>$$

There are N' doubles, $\left|\begin{matrix}2_i\bar{2}_i\\1_i\bar{1}_i\end{matrix}\right\rangle$ for $i = 1,2,\ldots,$ N', with equal coefficients $C$ in the expanded cluster. The operator $\hat{H}_{Lamb}$ has zero contribution in the minimal basis case, and the net contributions of operator $\hat{H}_{Hf}$ to equations (63), (66) and (69) also vanish. Therefore, the correlation energy is given by

$$E_{correl} = N'C(K_{12} + K_{12}^B). \quad (I.3)$$

The coefficients can be determined from (69) to obtain

$$C = (K_{12} + K_{12}^B)^{-1}\left[\Delta - [\Delta^2 + (K_{12} + K_{12}^B)^2]^{1/2}\right],$$

$$\Delta = \frac{1}{2}(E_2^* - E_2^0) = (\varepsilon_2 - \varepsilon_1) + \frac{1}{2}\left[(J_{11} + J_{11}^B) + (J_{22} + J_{22}^B)\right] - 2(J_{12} + J_{12}^B) + (K_{12} + K_{12}^B) \quad (I.4)$$

so that the correlation energy calculation is manifestly size-consistent,

$$E_{correl} = N'\left[\Delta - [\Delta^2 + (K_{12} + K_{12}^B)^2]^{1/2}\right]. \quad (I.5)$$

The molecule is strictly in the nonrelativistic limit as $Z = 1$. The overall effective nuclear charge is $Z_{eff}|e|$ where $Z_{eff}$ somewhat varies from $Z$. Henceforth in showing the orders $Z$ will be written in place of $Z_{eff}$. The normalized orbitals in the nonrelativistic limit are

$$\varphi_{(1 \text{ or } 2),(\uparrow \text{ or } \downarrow)} = (1\sigma \text{ or } 1\sigma^*)\chi_{\uparrow \text{ or } \downarrow},$$

$$\chi_{\uparrow \text{ or } \downarrow} = \begin{pmatrix}1\\0\end{pmatrix} \text{ or } \begin{pmatrix}0\\1\end{pmatrix}. \quad (I.6)$$

One obtains the expansion

$$(h_D)_{11} = \langle h_{nonrel}\rangle_{1\sigma} - \frac{\langle p^2\rangle_{1\sigma}^2}{8m^3c^2} + \frac{\pi\alpha^2\hbar^4}{2m^2e^2}\left[\{1\sigma(\mathbf{R}_1)\}^2 + \{1\sigma(\mathbf{R}_2)\}^2\right] + O(mc^2\alpha^5 Z^5), \quad (I.7)$$

the spin-orbit interaction being absent as the orbital angular momentum is zero. The operator $h_{nonrel}$ is the Schrödinger Hamiltonian operator for the H$_2$ molecule. A similar relation with $1\sigma^*$ in place of $1\sigma$ holds for $(h_D)_{22}$.

The two-electron integrals are found as

$$J_{11} = K_{11} = \left(1 - \frac{\langle p^2 \rangle_{u1}}{2m^2c^2}\right) J_{u1,u1} + 2J_{u1,l1} + O(mc^2\alpha^6 Z^6),$$

$$J_{12} = J_{u1,u2} + \delta J_{12},$$

$$\delta J_{12} = -\frac{\langle p^2 \rangle_{1\sigma} + \langle p^2 \rangle_{1\sigma^*}}{4m^2c^2} J_{1\sigma,1\sigma^*} + J_{u1,l2} + J_{l1,u2} + O(mc^2\alpha^6 Z^6), \quad (I.8)$$

$$K_{12} = K_{u1,u2} + \delta K_{12},$$

$$\delta K_{12} = -\frac{\langle p^2 \rangle_{1\sigma} + \langle p^2 \rangle_{1\sigma^*}}{4m^2c^2} K_{1\sigma,1\sigma^*} + K_{u1,l2} + K_{l1,u2} + O(mc^2\alpha^6 Z^6).$$

The Breit integrals are $J_{11}^B = \langle 11|B|11\rangle$, $J_{12}^B = \langle 12|B|12\rangle$, $J_{22}^B = \langle 22|B|22\rangle$ and $K_{12}^B = \langle 12|B|21\rangle$ where $B(1,2)$ is the Breit operator for the two electrons. These are of order $mc^2\alpha^4 Z^4$ and higher.

The nonrelativistic energies

$$E_{N,nonrel}^0 = N' E_{2,nonrel}^0,$$

$$E_{2,nonrel}^0 = 2\langle h_{nonrel}\rangle_{1\sigma} + J_{1\sigma,1\sigma},$$

$$\varepsilon_{1,nonrel} = \langle h_{nonrel}\rangle_{1\sigma} + J_{1\sigma,1\sigma}, \quad (I.9)$$

$$\varepsilon_{2,nonrel} = \langle h_{nonrel}\rangle_{1\sigma^*} + 2J_{1\sigma,1\sigma^*} - K_{1\sigma,1\sigma^*},$$

$$E_{2,nonrel}^* = 2\langle h_{nonrel}\rangle_{1\sigma^*} + J_{1\sigma^*,1\sigma^*},$$

are supplemented by the familiar relativistic corrections

$$\frac{\delta E_{N,rel}^0}{N'} = \delta E_{2,rel}^0 = -\frac{\langle p^2 \rangle_{1\sigma}^2}{4m^3c^2} + \frac{2\pi\alpha^2 Z\hbar^4}{m^2 e^2}[1\sigma(\mathbf{R}_1)]^2 - \frac{\langle p^2 \rangle_{1\sigma}}{2m^2c^2} J_{1\sigma,1\sigma} + 2J_{u1,l1} + O(mc^2\alpha^5 Z^5),$$

$$\delta\varepsilon_{1,rel} = -\frac{\langle p^2 \rangle_{1\sigma}^2}{8m^3c^2} + \frac{\pi\alpha^2 Z\hbar^4}{m^2 e^2}[1\sigma(\mathbf{R}_1)]^2 - \frac{\langle p^2 \rangle_{1\sigma}}{2m^2c^2} J_{1\sigma,1\sigma} + 2J_{u1,l1} + O(mc^2\alpha^5 Z^5),$$

$$\delta\varepsilon_{2,rel} = -\frac{\langle p^2 \rangle_{1\sigma^*}^2}{8m^3c^2} + \frac{\pi\alpha^2 Z\hbar^4}{m^2 e^2}[1\sigma^*(\mathbf{R}_1)]^2 + 2\delta J_{12} - \delta K_{12} + O(mc^2\alpha^5 Z^5),$$

$$\delta E_{2,rel}^* = -\frac{\langle p^2 \rangle_{1\sigma^*}^2}{4m^3c^2} + \frac{2\pi\alpha^2 Z\hbar^4}{m^2 e^2}[1\sigma^*(\mathbf{R}_1)]^2 - \frac{\langle p^2 \rangle_{1\sigma}}{2m^2c^2} J_{1\sigma^*,1\sigma^*} + 2J_{u2,l2} + O(mc^2\alpha^5 Z^5).$$

(I.10)

The $p^2/m^3c^2$ term is the kinetic correction, the contact term is the Darwin interaction that arises from "zitterbewegung" – rapid oscillatory motion of the electron within the nucleus, and additional corrections are obtained from the two-electron interaction. These are accompanied by the QED corrections

$$\frac{\delta E_{N,QED}^0}{N'} = \delta E_{2,QED}^0 = J_{11}^B,$$

$$\delta\varepsilon_{1,QED} = J_{11}^B, \quad (I.11)$$

$$\delta\varepsilon_{2,QED} = 2J_{12}^B - K_{12}^B,$$

$$\delta E_{2,QED}^* = J_{22}^B.$$

Furthermore, the hyperfine splitting of orbitals is calculated using the proton spinors $u_{P\uparrow} = \begin{pmatrix} 1 \\ 0 \end{pmatrix}$, $l_{P\uparrow} = \mathbf{0}$ and $u_{P\downarrow} = \begin{pmatrix} 0 \\ 1 \end{pmatrix}$, $l_{P\downarrow} = \mathbf{0}$ such that $\beta_P$ may be replaced by the unit matrix of rank 4. The total nuclear spin states are written in this notation as

$$\text{Singlet}: \quad |I=0, I_z=0\rangle = 2^{-1/2}(|I_{1\uparrow}I_{2\downarrow}\rangle - |I_{1\downarrow}I_{2\uparrow}\rangle);$$
$$\text{Triplet}: \quad |1,1\rangle = |I_{1\uparrow}I_{2\uparrow}\rangle; \quad |1,0\rangle = 2^{-1/2}(|I_{1\uparrow}I_{2\downarrow}\rangle + |I_{1\downarrow}I_{2\uparrow}\rangle); \quad |1,-1\rangle = |I_{1\downarrow}I_{2\downarrow}\rangle.$$
(I.12)

The hyperfine interaction contributes to the orbital energies when the nuclear spin states are $|1,1\rangle$ and $|1,-1\rangle$. To order $\alpha^2 Z m_e/M_P$, contributions to the $1\sigma$ spinor energies are

$$\varepsilon_{1,(00)}^{hf} = \varepsilon_{\bar{1},(00)}^{hf} = \varepsilon_{1,(10)}^{hf} = \varepsilon_{\bar{1},(10)}^{hf} = 0,$$ (I.13)

$$\varepsilon_{1,(11)}^{hf} = -\varepsilon_{\bar{1},(11)}^{hf} = -\varepsilon_{1,(1\bar{1})}^{hf} = \varepsilon_{\bar{1},(1\bar{1})}^{hf} = \Delta_{hf}(1\sigma),$$

$$\Delta_{hf}(1\sigma) = -\frac{e^2 \hbar^2 g_e g_P}{4 m M_P c^2} \Big[ \langle 1\sigma | \frac{1}{|\mathbf{r}-\mathbf{R}_1|^3} + \frac{1}{|\mathbf{r}-\mathbf{R}_2|^3} | 1\sigma \rangle$$ (I.14)

$$-3\langle 1\sigma | \frac{(z-Z_1)^2}{|\mathbf{r}-\mathbf{R}_1|^5} + \frac{(z-Z_2)^2}{|\mathbf{r}-\mathbf{R}_2|^5} | 1\sigma \rangle - \frac{8\pi}{3}[|1\sigma(\mathbf{r}=\mathbf{R}_1)|^2 + |1\sigma(\mathbf{r}=\mathbf{R}_2)|^2] \Big],$$

where $|\Delta_{hf}(1\sigma)|$ is the hyperfine splitting of each electronic spin-orbital. Similar expressions are obtained for hyperfine corrections to $\varepsilon_2$. However, the hyperfine corrections to orbitals do not contribute to total energy in the ground state of a closed shell molecule, as the hyperfine energies of two different electron spins cancel each other.

Finally, the ground state correlation energy exhibits the trends (through order $mc^2\alpha^4 Z^4$):

$$\delta\Delta_{rel} = \Delta_{rel} - \Delta_{nonrel} = \frac{1}{2}(\delta E_{2,rel}^* - \delta E_{2,rel}^0),$$ (I.15)

$$\frac{\delta E_{correl,rel}}{N'} = \delta\Delta_{rel} - \frac{\Delta_{nonrel}\delta\Delta_{rel} + K_{12}\delta K_{12}}{(\Delta_{nonrel}^2 + K_{12}^2)^{1/2}}$$
$$= \frac{N'^{-1}E_{correl,nonrel}\delta\Delta_{rel} + K_{12}\delta K_{12}}{N'^{-1}E_{correl,nonrel} - \Delta_{nonrel}},$$ (I.16)

$$\delta\Delta_{QED} = \frac{1}{2}(J_{22}^B - J_{11}^B),$$ (I.17)

$$\frac{\delta E_{correl,QED}}{N'} = \delta\Delta_{QED} - \frac{\Delta_{nonrel}\delta\Delta_{QED} + K_{12}K_{12}^B}{(\Delta_{nonrel}^2 + K_{12}^2)^{1/2}}$$
$$= \frac{N'^{-1}E_{correl,nonrel}\delta\Delta_{QED} + K_{12}K_{12}^B}{N'^{-1}E_{correl,nonrel} - \Delta_{nonrel}}.$$ (I.18)

After some calculations, the pair energy values are found:

$$\frac{\delta E_{1-pair}}{N'} = \frac{1}{mc^2}\Big[|<11|21'>|^2 + |<11|21'>|^2\Big] \sim O(mc^2\alpha^6 Z^6),$$
$$\frac{\delta E_{2-pair}}{N'} = \frac{1}{4mc^2}\Big[|K_{11'}|^2 + |K_{22'}|^2 + 2|<11|1'2'>|^2\Big] \sim O(mc^2\alpha^8 Z^8).$$ (I.19)

The 2-pair correction is negligibly smaller than the 1-pair term. Even the 1-pair term is smaller than the relativistic and other QED corrections in absolute magnitude by an order of 2 in fine structure constant. The size consistency is obvious at every step of calculation – not only at the known nonrelativistic level but also in relativistic corrections, radiative effects, relativistic correlation energy and pair energies.